\documentclass[10pt]{article}
\usepackage{graphicx}
\usepackage[cp1251]{inputenc}

\begin{document}

\twocolumn [\noindent{\small\it ISSN 1063--7737, Astronomy
Letters, 2010, Vol. 36, No. 7, pp. 457--466. \copyright Pleiades
Publishing, Inc., 2010. Original Russian Text \copyright A.T.
Bajkova, A.B. Pushkarev, 2010, published in Pis'ma v
Astronomicheski$\hat{i}$ Zhurnal, 2010, Vol. 36, No. 7, pp.
483--493.}

\vskip -4mm

\begin{tabular}{llllllllllllllllllllllllllllllllllllllllllllllll}
 & & & & & & & & & & & & & & & & & & & & & & & & & & & & & & & & & & & & & & & \\
\hline \hline
\end{tabular}

\vskip 1.5cm

\centerline{\large\bf Multifrequency Method for Mapping Active
Galactic Nuclei} \centerline{\large\bf with Allowance for the
Frequency-Dependent Image Shift}

\bigskip

\centerline{\bf  A. T. Bajkova$^1$ and A. B. Pushkarev$^{1,2,3}$}

\medskip

\centerline{\it $^1$Pulkovo Astronomical Observatory of RAS,
Pulkovskoe sh. 65, St. Petersburg, 196140 Russia}

\centerline{e-mail: {\it bajkova@gao.spb.ru}}

\centerline{\it $^2$Max Planck Institut f$\ddot{u}$r
Radioastronomie, Auf dem H$\ddot{u}$gel 69, D-53121 Bonn, Germany}

\centerline{\it $^3$Crimean Astrophysical Observatory, pos.
Nauchnyi, Crimea, 98409 Ukraine}

\medskip

\centerline{\small Received August 26, 2009}

\vskip 1.5cm

{\bf Abstract} --- {\small We consider the problem of
multifrequency VLBA image synthesis and spectral-index mapping for
active galactic nuclei related to the necessity of taking into
account the frequency-dependent image shift. We describe our
generalized multifrequency synthesis algorithm with a spectral
correction based on the maximum entropy method. The results of our
processing of multifrequency VLBI data for the radio sources
J2202+4216, J0336+3218, and J1419+5423 are presented.  }

\medskip

\noindent DOI: 10.1134/S1063773710070017

\medskip


\vskip 1cm
]

\centerline{INTRODUCTION }

\medskip

Active galactic nuclei (AGN) (quasars, blazars, radio galaxies,
and Seyfert galaxies) are among the most enigmatic objects in our
Universe. The AGN phenomenon consists in colossal energy release
(up to $10^{46-47}$ erg s$^{-1}$) from relatively small spatial
scales ($R<1$ pc). Several theoretical models have been proposed
to explain the nature of AGN, with the model of a supermassive (up
to $10^{8\pm2}$M$_\odot$) black hole at the galactic center onto
which matter accretes having gained the widest acceptance. As a
result of accretion, a colossal energy emitted by AGN is
generated; the flows of magnetized ultra-relativistic plasma are
ejected along the rotation axis of the accreting disk with a speed
close to the speed of light to produce jets observed at parsec and
kiloparsec distances from the "central engine" (Marscher and
Erstadt 2007). The signatures of AGN activity include a
non-thermal spectrum attributable mainly to synchrotron radiation
generated by the motion of electrons with relativistic velocities
in a magnetic field; strong variability; the presence of broad
emission lines in optical spectra; the presence of relativistic
jets; spectral and polarization peculiarities.

At present, very-long-baseline interferometry (VLBI) is the most
powerful tool for studying the morphological structures and
kinematic, polarization, and spectral characteristics of AGN; it
allows objects to be imaged with a very high angular resolution
reaching fractions of a milliarcsecond (mas). One of the topical
problems of VLBI mapping for AGN is multifrequency image
synthesis. Our interest in this method is mainly related to the
peculiar geometry of the future high-orbit ground.space
"Radioastron" radio interferometer (Kardashev 1997), which will
provide an ultrahigh resolution (microarcseconds), on the one
hand, and poor aperture filling, on the other hand (Bajkova 2005).

Multifrequency synthesis in VLBI suggests mapping AGN at several
frequencies simultaneously to improve the instrument aperture
filling. This is possible, because the interferometer baselines
are measured in wavelengths of the emission being received. The
problem of multifrequency synthesis is complicated by the fact
that the brightness of a radio source generally depends on the
frequency and to avoid undesirable artifacts in the image, a
spectral correction should be made at the stage of its
deconvolution.

Conway et al. (1990), Conway (1991), Sault and Wieringa (1994),
and Sault and Conway (1999) investigated the influence of spectral
effects on the image and developed the methods of their
correction. These authors showed that if narrow frequency bands,
up to $\pm$12.5\% of the reference frequency, are used, then the
effects of the spectral dependence of the brightness of a radio
source can be neglected in principle, because they are small and
can be compensated at the calibration or self-calibration stage.
However, in the case of broader bands, for example, $\pm$30\%, the
spectral effects should be corrected.

The algorithm of a linear spectral correction based on the CLEAN
method (H$\ddot{o}$ogbom 1974) and called "double deconvolution"
(Conway et al. 1990) is the best-studied one. In this algorithm,
the "dirty" image is first deconvolved with an ordinary "dirty"
beam and the residual map is then deconvolved with the beam
responsible for the first-order spectral term. The development of
this method consisting in simultaneous reconstruction of the
sought-for image and the map of the spectral term was proposed by
Sault and Wieringa (1994). The vector relaxation algorithm
developed by Likhachev et al. (2006) may be considered a
generalized CLEAN-deconvolution method that includes the spectral
terms of any order.

As the study by Bajkova (2008) showed, applying Shannon's maximum
entropy method (MEM) (Frieden 1972; Skilling and Bryan 1984;
Narayan and Nityananda 1986; Frieden and Bajkova 1994) allows
tangible progress to be achieved in solving the problem of
multifrequency synthesis owing to the possibility of a simple
allowance for the spectral terms of any order. This, in turn,
allows the range of synthesized frequencies to be extended
significantly (up to 90\% of the reference frequency).

However, the multifrequency synthesis algorithms based on both
CLEAN and MEM deconvolution that we discuss here ca n be directly
applied only to those radio sources for which no
frequency-dependent image shift is observed. Otherwise, as was
shown by Croke and Gabuzda (2008), an additional operation to
align images at different frequencies should be performed to
obtain the proper results of a multifrequency data analysis.

In this paper, our goal is to deduce our multifrequency synthesis
algorithm based on MEM and to show how important the procedure for
precorrecting the frequency-dependent image shift is in
implementing multifrequency synthesis.

The paper is structured as follows. The frequency dependence of
the image of a radio source is described in the next section.
Subsequently, frequency dependent constraints on the visibility
function are derived. The developed multifrequency synthesis
algorithm with a frequency correction based on MEM is briefly
deduced.We discuss the problem of aligning frequency-dependent
images to properly construct the spectral-index distribution and,
as an example, present the results of applying the developed
method to multifrequency VLBI data for three radio sources:
J2202+4216, J0336+3218, and J1419+5423.

\medskip

\centerline{THE FREQUENCY DEPENDENCE} \centerline{OF THE AGN RADIO
BRIGHTNESS}

\medskip

The dependence of the intensity of a radio source on frequency
$\nu$ in the model of synchrotron radiation is given by (Conway et
al. 1990)
\begin{equation}
I(\nu)=I(\nu_0)\left(\frac{\nu}{\nu_0}\right)^\alpha,
\end{equation}
where $I(\nu_0)$ is the intensity of the radiation at the
reference frequency $\nu_0$; and $\alpha$ is the spectral index.
Below, to simplify the writing, we will everywhere set
$I_0=I(\nu_0)$.

Retaining the first $Q$ terms in the Taylor expansion of (1) at
point $\nu_0$, we can write the following approximate equality:
\begin{equation}
I(\nu)\approx I_0+\sum_{q=1}^{Q-1}I_q
\left(\frac{\nu-\nu_0}{\nu_0}\right)^q,
\end{equation}
 where
$$I_q=I_0\frac{\alpha(\alpha-1)\cdots[\alpha-(q-1)]}{q!}.$$

In accordance with Eq. (2), for each point (l,m) of the source's
two-dimensional ($N\times N$) brightness distribution (image,
map), we have
\begin{equation}
I(l,m)\approx I_0(l,m)+\sum_{q=1}^{Q-1}I_q(l,m)
\left(\frac{\nu-\nu_0}{\nu_0}\right)^q,
\end{equation}
where $l,m=1,...,N$.

Thus, the derived brightness distribution over the source (3) is
the sum of the brightness distribution at the reference frequency
$\nu_0$ and the spectral terms, with the qth-order spectral map
depending on the spectral-index distribution over the source as
follows:
\begin{eqnarray}
I_q(l,m)=I_0(l,m)\\
\times\frac{\alpha(l,m)[\alpha(l,m)-1]\cdots[\alpha(l,m)-(q-1)]}{q!}.\nonumber
\end{eqnarray}
Of greatest interest is the first-order spectral map
\begin{equation}
I_1(l,m)=I_0(l,m)\alpha(l,m).
\end{equation}

The spectral-index distribution over the source can be estimated
from Eq. (5):
\begin{equation}
\alpha(l,m)=I_1(l,m)/I_0(l,m).
\end{equation}

\medskip

\centerline{CONSTRAINTS ON THE VISIBILITY} \centerline{FUNCTION}

\medskip

The complex visibility function is the Fourier transform of the
intensity distribution over the source that satisfies the spectral
dependence (1) at each point of the map (l,m). Given the finite
number of terms in the Taylor expansion (3), the constraints on
the visibility function can be written as
\begin{eqnarray}
V_{u_\nu,v_\nu}= {\bf F}\{I(l,m)\}\times {\bf D}_{u_\nu,v_\nu}\\
\approx \sum_{q=0}^{Q-1} {\bf F}\left\{I_q(l,m)
\left(\frac{\nu-\nu_0}{\nu_0}\right)^q\right\}\times {\bf
D}_{u_\nu,v_\nu},\nonumber
\end{eqnarray}
where {\bf F}  denotes the Fourier transform and {\bf D} denotes
the transfer function, which is the $\delta$-function of $u$ and
$v$ for each measurement of the visibility function; different
sets of $\delta$-functions correspond to different frequencies
$\nu$, as suggested by the indices of $u$ and $v$.

Let us rewrite Eq. (7) for the real and imaginary parts of the
visibility function
$V_{u_\nu,v_\nu}=A_{u_\nu,v_\nu}+jB_{u_\nu,v_\nu}$ by taking into
account the measurement errors as
\begin{equation}
\sum_{q=0}^{Q-1}\sum_{l,m} I_q(l,m)
a^{lm}_{u_\nu,v_\nu}\left(\frac{\nu-\nu_0}{\nu_0}\right)^q+\eta^{re}_{u_\nu,v_\nu}=A_{u_\nu,v_\nu},
\end{equation}
\begin{equation}
\sum_{q=0}^{Q-1}\sum_{l,m} I_q(l,m)
b^{lm}_{u_\nu,v_\nu}\left(\frac{\nu-\nu_0}{\nu_0}\right)^q+\eta^{im}_{u_\nu,v_\nu}=B_{u_\nu,v_\nu},
\end{equation}
where $a^{lm}_{u_\nu,v_\nu}$ and $b^{lm}_{u_\nu,v_\nu}$ are the
constant coefficients (cosines and sines) that correspond to the
Fourier transform;  $\eta^{re}_{u_\nu,v_\nu}$ and
$\eta^{im}_{u_\nu,v_\nu}$ are the real and imaginary parts of the
instrumental additive noise distributed normally with a zero mean
and a known dispersion $\sigma_{u_\nu,v_\nu}$.

\medskip

\centerline{THE MULTIFREQUENCY SYNTHESIS} \centerline{ALGORITHM}

\medskip

In this case, the distributions $I_q(l,m)$, $q=0,...,Q-1;~~
l,m=1,...,N$, and the measurement errors of the visibility
function $\eta^{re}_{u_\nu,v_\nu},~\eta^{im}_{u_\nu,v_\nu}$ are
unknown.  Note that although the brightness distribution over the
source is described by an nonnegative function, the spectral maps
of arbitrary order (4) can generally take on both positive and
negative values, because, in particular, the spectral-index
distribution over the source is an alternating one.  Since the
logarithm of a negative value is not defined on the set of real
numbers, we will form a functional in which the spectral maps
appear in absolute value to find the solutions for
$I_q(l,m),~q=0,...,Q-1$:
\begin{eqnarray}
{\bf E}=\{\sum_{l,m} I_0(l,m)\ln [I_0(l,m)]\\
+\sum_{q=1}^{Q-1} \sum_{l,m} |I_q(l,m)|\ln
[|I_q(l,m)|]\}\nonumber\\
+\rho\sum_{u_\nu,v_\nu}\frac{(\eta^{re}_{u_\nu,v_\nu})^2+(\eta^{im}_{u_\nu,v_\nu})^2}{\sigma^2_{u_\nu,v_\nu}},\nonumber
\end{eqnarray}
\begin{equation}
I_0(l,m)\ge 0,
\end{equation}
 where $\rho$ is a positive weighting factor.

Minimizing functional (10) with constraints (8), (9), and (11)
constitutes the essence of the MEM-based multifrequency synthesis
algorithm.

The last term in (10) is an estimate of the disagreement of the
reconstructed spectrum with the measured data according to the
$\chi^2$ test and may be considered as an additional stabilizing
term. In this case, the influence of this term on the resolution
of the reconstruction algorithm should be kept in mind.

For a practical solution of the reconstruction problem, let us
pass to the generalized maximum entropy method described in detail
by Bajkova (1992, 1993, 2005, 2008) and Frieden and Bajkova
(1994).It consists in the substitution
\begin{equation}
I_q(l,m)=I_q^+(l,m)-I_q^-(l,m),
\end{equation}
where the superscripts + and . denote the positive and negative
parts of the function, respectively, and the following
modification of functional (10):
\begin{eqnarray}
{\bf E}=\sum_{l,m} I_0(l,m)\ln [a I_0(l,m)]\\
+\sum_{q=1}^{Q-1} \sum_{l,m} \{I_q^+(l,m)\ln [a I_q^+(l,m)]\nonumber\\
+I_q^-(l,m)\ln [a I_q^-(l,m)]\}\nonumber\\
+\rho\sum_{u_\nu,v_\nu}\frac{(\eta^{re}_{u_\nu,v_\nu})^2+(\eta^{im}_{u_\nu,v_\nu})^2}{\sigma^2_{u_\nu,v_\nu}},\nonumber
\end{eqnarray}
\begin{equation}
I_0(l,m)\ge 0,~~~~~I_q^+(l,m)\ge 0,~~~~~I_q^-(l,m)\ge 0,
\end{equation}
where $a\gg 1$ s the parameter responsible for the accuracy of
separating the positive,  $I_q^+(l,m)$, and negative, I.q (l,m),
parts of the solution for $I_q(l,m)$ (Bajkova 1992).

The linear constraints (8) and (9) on the measured visibility
function will be rewritten accordingly:
\begin{eqnarray}
{\bf R_A}=\sum_{l,m} I_0(l,m) a^{lm}_{u_\nu,v_\nu}\\
+\sum_{q=1}^{Q-1}\sum_{l,m} [I_q^+(l,m)-I_q^-(l,m)]
a^{lm}_{u_\nu,v_\nu}\nonumber\\
\times\left(\frac{\nu-\nu_0}{\nu_0}\right)^q+\eta^{re}_{u_\nu,v_\nu}=A_{u_\nu,v_\nu},\nonumber
\end{eqnarray}
\begin{eqnarray}
{\bf R_B}=\sum_{l,m} I_0(l,m)
b^{lm}_{u_\nu,v_\nu}\\
+\sum_{q=1}^{Q-1}\sum_{l,m} [I_q^+(l,m)-I_q^-(l,m)]
b^{lm}_{u_\nu,v_\nu}\nonumber\\
\times\left(\frac{\nu-\nu_0}{\nu_0}\right)^q+\eta^{im}_{u_\nu,v_\nu}=B_{u_\nu,v_\nu},\nonumber
\end{eqnarray}
where ${\bf R_A}$ and ${\bf R_B}$ denote the left-hand sides of
the equations.

Thus, reconstructing the image $I_0(l,m)$ requires optimizing
functional (13), i.e., finding
\begin{equation}
\min {\bf E}
\end{equation}
under constraints (14)-(16) for all unknown
$I_0(l,m),~I_q^{+(-)}(l,m),~q=1,...,Q-1,~l,m=1,...,N $ and
$\eta^{re}_{u_\nu,v_\nu},~\eta^{im}_{u_\nu,v_\nu}$. Note that
requirements (14) can be omitted due to the peculiarity of the
entropy solution, which is purely positive, and then only the
linear constraints (15) and (16) on the measured complex
visibility function remain. A detailed algorithm for numerical
implementation of the proposed multifrequency synthesis method is
given in Bajkova (2008).

\sloppy  The advantage of the proposed multifrequency synthesis
algorithm is that the spectral terms of any order can be easily
taken into account in functional (13) being minimized. This allows
the spectral correction of images to be made both in a wide
frequency range and for large spectral indices. The results of
careful modeling of the method and estimates for the quality of
the spectral correction of images depending on the size of the
frequency band being synthesized, the spectral index, and the
quality of the measurements of the visibility function are
presented in Bajkova (2008).

\medskip

\centerline{ THE PROBLEM OF SPECTRAL-INDEX} \centerline{MAPPING }

\medskip

 One of the most important sources of
information about the physical conditions in the radio-emission
regions of AGN is the spectral-index distribution over the source.
The core region is usually characterized by a large optical depth
and an almost flat or inverted spectrum, while the jets are
optically thin with respect to synchrotron radiation and have
steeper spectra (Pushkarev and Kovalev 2009; Croke and
Gabuzda2008).

The spectral-index distribution over the source can be constructed
by various methods. The traditional method suggests: (1) the
formation of images at two separate frequencies ($\nu_1$ and
$\nu_2$), with the solutions of the deconvolution problem (CLEAN
or MEM) being convolved with the same clean beam corresponding to
the lower observing frequency; (2) determining the two-dimensional
spectral-index distribution over the source from Eq. (1).
Obviously, this sequence of operations is legitimate only when the
positions of the VLBI cores of sources (not to be confused with
the physical core of the source that is undetectable due to
absorption effects) are frequency-independent.

The image reconstruction using the iterative selfcalibration
procedure is known (Thompson et al. 2003) to lead to the loss of
information about the absolute position of the source in the sky:
during the phase self-calibration, the centroid of the object is
placed at the phase center of the map with coordinates (0,0).
However, since most of the radio-loud AGN are characterized by a
dominant compact core (Kovalev et al. 2005; Lee et al. 2008;
Pushkarev and Kovalev 2009), the VLBI core of the source coincides
with the peak radio brightness of the source in an overwhelming
majority of cases.

Nevertheless, the standard theory of extragalactic radio sources
(Blandford and Konigl 1979) predicts a frequency-dependent VLBI
core shift due to opacity effects in the source's core region.
Synchrotron self-absorption takes place in an ultra-compact region
near the "central engine" of AGN, whose mechanism is most
efficient at low frequencies. As a result, the maximum brightness
manifests itself farther from the core along the jet axis at lower
frequencies. This theoretical prediction was confirmed by
observations: the frequency-dependent shift in the core position
was measured for several quasars by Lobanov (1998). In the
literature, this phenomenon is actively debated from the viewpoint
of the accuracy of astrometric measurements (Charlot 2002; Boboltz
2006; Kovalev et al. 2008).

It thus follows that the multifrequency data analysis must be
preceded by the alignment of images at different frequencies. This
can be achieved in three ways: (1) performing VLBI observations of
the objects under study together with reference sources; (2)
finding the parameters of the shift of one image relative to the
other by aligning compact features of the optically thin jet,
which are not subjected to absorption effects to the same extent
as in the source's core (Paragi et al. 2000; Kovalev et al.
2008a); (3) finding the shift parameters using a cross-correlation
analysis (Croke and Gabuzda 2008). Being laborious from the
viewpoint of performing observations and their subsequent
reduction, the first method gives no significant advantage in
determining the shift; therefore, the second and/or third methods
are used more often.

Recall that the alignment procedure implemented by shifting one
image relative to the other is equivalent to the phase correction
of the spectrum (or visibility function) of the image being
shifted relative to the fixed one. The necessity of precorrecting
the data for the source's visibility function at different
frequencies makes the direct use of the multifrequency synthesis
algorithm described above problematic, because the frequency
dependence of the core shift is not known in advance. It can be
determined by forming the images at each frequency and determining
the corresponding shifts. As was shown by Kovalev et al. (2008b)
and O'Sullivan and Gabuzda (2009), the frequency dependence of the
VLBI core position is well fitted by a hyperbolic dependence of
the form $r\propto\nu^{-1}$. Thus, our multifrequency synthesis
procedure can be used after allowance for the shifts in the
positions of the VLBI cores at different frequencies and their
coordinates relative to the phase center and applying the
corresponding frequency-dependent phase corrections to the
visibility function.

\medskip

\centerline{ RESULTS OF OUR PROCESSING} \centerline{OF REAL DATA }

\medskip

We will present the results of applying the developed
multifrequency image synthesis algorithm to three representatives
of AGN: J2202+4216, J0336+3218, and J1419+5423. All these sources
have a fairly complex structure that includes an optically thick
VLBI core and an optically thin extended jet, which must manifest
itself in the spectral-index distribution.

The observations performed with the VLBA antennas and several
antennas of the global VLBI network were taken from the NRAO
archive \footnote{\tt http://archive.nrao.edu}. The da ta on the
visibility function at various frequencies were obtained
simultaneously in "snapshot" mode. The data were calibrated using
standard procedures from the AIPS package. The images were formed
using procedures based on MEM and its generalizations within the
framework of the Pulkovo "VLBImager" software package.

The object names, dates and frequencies of observations, and
parameters of the synthesized maps and the smoothing "clean" beam
that determines the system's resolution are given in Table 1. The
parameters of the maps synthesized from the observational data at
different frequencies and the parameters of the
frequency-dependent image shift found by aligning compact features
of the optically thin jet are given in Table 2. As expected, the
direction of the shift coincides with the jet direction in all
cases. The images of the sources obtained from both
single-frequency and multi-frequency data and the two-dimensional
spectral-index distributions over the source are shown in Figs.
1--6.

Let us analyze the results obtained.

 (1) {\bf J2202+4216.} Figure 1
presents the images of the radio source obtained separately at 2.3
and 8.6 GHz. The source consists of an optically thick core and an
optically thin jet directed southward of the core. The apparent
extent of the jet reaches 36 and 8 mas at the low and high
frequencies, respectively.

Figure 2 shows the intensity map and the map of the spectral-index
distribution over the source synthesized at a reference frequency
of 5.5 GHz from the data at 2.3 and 8.6GHz without any allowance
for the image shift found. As can be seen, the spectral-index map
does not correspond to the physical meaning of an optically thick
core and an optically thin jet: there are segments with negative
spectral indices in the core region and with positive spectral
indices in the jet region. The source's image is not compact
either: the core shape is severely distorted.

Figure 3 shows the result of two-frequency synthesis with the
correction of the shift found by aligning the peak values of the
single-frequency images. As we see, we managed to improve
noticeably the previous result, but it is not yet quite proper,
because the frequency-dependent shift of the peak values does not
reflect the real shift of the images relative to each other.

Only allowance for the real shift found by aligning features of
the optically thin jet, which is virtually unaffected by
absorption effects, yielded the proper result shown in Fig. 4. As
we see, as a result of the synthesis, we managed to reconstruct a
more extended jet structure (up to 14 mas) than in the case of
using only the high-frequency data but with the same high angular
resolution. The spectral-index map adequately reflects the
physical characteristics of the regions of the optically thick
compact VLBI core and the optically thin extended jet. We see a
fairly regular structure with smooth transitions between segments
of different intensities along the entire source. The presented
result agrees well with that obtained by Croke and Gabuzda (2008).

(2) {\bf J0336+3218.} Figure 5 presents both the single-frequency
images of the radio source (upper panel) obtained separately at
2.3 and 8.6GHz and the maps of the intensity and spectral-index
distributions over the source obtained through two-frequency
synthesis at a reference frequency of 5.5 GHz (lower panel). The
source consists of an optically thick core and an optically thin
jet extending roughly southwestward of the core. The jet extent
reaches about 27 and 10 mas at the low and high frequencies,
respectively.

As a result of the two-frequency synthesis with allowance made for
the shift found, we managed to obtain the proper maps of the
intensity and spectral-index distributions over the source. As we
see from Fig. 5c, we managed to reconstruct a more extended jet
structure (up to 15 mas) than in the case of using only the
high-frequency data, with the synthesized map having the angular
resolution corresponding to the high-frequency data. The
spectral-index map adequately reflects the physical
characteristics of the regions of the optically thick compact VLBI
core, where the spectral index is positive, and the optically thin
extended jet, where the spectral index is negative. We see that
the spectral-index distribution reflects the spatial structure of
the jet that consists of several compact features, showing smooth
transitions between segments of different intensities.

(3) {\bf J1419+5423.} This source is represented by observational
data simultaneously at three frequencies: 5, 8.4, and 15.3 GHz
(see Table 1). As a result, we have the possibility of
three-frequency synthesis. The intensity maps obtained at
individual frequencies are shown in the upper panel of Fig. 6. As
can be seen, the source consists of a compact core and an extended
jet extending southwestward of the core to a distance of about 25
mas. The compact features of the jet at distances of about 5 and
18 mas from the core are clearly seen on the maps constructed from
the data at 5 and 8.4 GHz. The map constructed from the data at
15.3 GHz has a higher angular resolution and clearly shows the
inner component of the jet at a distance of about 1 mas from the
core, but, at the same time, the component at 5 mas is barely seen
and the component at 18 mas is completely invisible.

The lower panel of Fig. 6 shows the results of our three-frequency
synthesis with a reference frequency of 8.4 GHz obtained by taking
into account the relative shift of the single-frequency images
found. As can be seen from our intensity map, as a result of the
three-frequency synthesis, we managed to reconstruct all jet
components, both the component closest to the core and the farther
ones. Thus, just as for the two previous sources, both a high
resolution of the source's components and a high sensitivity to
the extended jet component can be achieved through the use of both
low-frequency and high-frequency measurements.

Figure 6e shows the spectral-index map corresponding to the
resolution of the synthesized image (Fig. 6d), from which the
spectral-index distribution in the region of the resolved inner
jet component can be judged. Figure 6f shows the map obtained with
a low resolution corresponding to the measurements at the lowest
observing frequency (in the C band), from which the larger-scale
spectral-index distribution over the source can be judged. Thus,
it is pertinent to note that, in contrast to the traditional
method dealing with single-frequency images convolved with a wide
beam corresponding to the lowest observing frequency,
multifrequency synthesis allows the spectral index to be mapped
with a high spatial resolution corresponding to that of the
synthesized image.

\medskip

\centerline{CONCLUSIONS }

\medskip

We developed and tested an efficient multifrequency image
synthesis algorithm with the correction of the frequency
dependence of the radio brightness of a source. The algorithm is
based on the maximum entropy method; it allow one to take into
account the spectral terms of any order and to map the spectral
index, which is of great importance in investigating the physical
characteristics of AGN.

We showed how important the allowance for the frequency-dependent
image shift is in applying the multifrequency synthesis algorithm.
Our conclusions are based on the results of processing
multifrequency VLBA data for the radio sources J2202+4216,
J0336+3218, and J1419+5423 with a fairly complex extended jet
structure, which also manifests itself in the spectral-index
distribution over the source.

Analysis of the results obtained shows that multifrequency
synthesis is an efficient method for improving the mapping
quality; low-frequency data allow the extended structure of a
source to be reconstructed more completely, while high-frequency
data allow a high spatial resolution to be achieved.

As an additional advantage of the multifrequency synthesis
algorithm, it should also be emphasized that the spectral-index
distribution can be mapped with a high resolution corresponding to
that of the synthesized image of the source, while the traditional
spectral-index mapping method deals with images whose resolution
is determined by the lowest observing frequency.

\medskip

\centerline{ACKNOWLEDGMENTS}

\medskip

This work was supported by the "Origin and Evolution of Stars and
Galaxies" Program of the Presidium of the Russian Academy of
Sciences and the Program of State Support for Leading Scientific
Schools of the Russian Federation (grant no. NSh-6110.2008.2
"Multi-wavelength Astrophysical Research").

\clearpage \onecolumn

\centerline{{\bf Table 1.} Parameters of single-frequency maps}
\begin{center}\begin{small}
\begin{tabular}{|c|c|c|c|c|c|}\hline
           & Date of    & Observing    &                  &                       &  \\
Object     & observa- & frequency, & Frequency range  & Beam, mas $\times$ mas,[$^\circ$]  & Peak flux   \\
           &  tions          & GHz          &          &                              & Jy/beam  \\
\hline
J2202+4216 & 10.05.1999 & 2.292      & S           & $3.40\times 2.50,$ $-14^\circ$   & 1.58    \\
           &            & 8.646      & X           & $1.00\times 0.70, -14^o$           & 1.05    \\
\hline
J0336+3218 & 05.07.2001 & 2.302      & S           & $4.00\times 2.43,$ $-12^\circ$    & 0.83    \\
           &            & 8.646      & X           & $1.04\times 0.67,$ $-13^\circ$ & 0.76    \\
\hline
J1419+5423 & 09.02.1997 & 4.971      & C           & $2.18\times 1.93,$ $84^\circ$  & 0.58    \\
           &            & 8.405      & X           & $1.31\times 1.14,$ $85^\circ$  & 0.57    \\
           &            & 15.349     & U           & $0.70\times 0.62,$ $75^\circ$  & 0.56    \\
\hline
\end{tabular}\end{small}
\end{center}

\vskip 5cm

\centerline{{\bf  Table 2.} Parameters of multifrequency maps}
\begin{center}\begin{small}
\begin{tabular}{|c|c|c|c|c|c|c|}\hline
           &  Synthesized & Reference  &                        &                      &                     & \\
 Object    & frequen-     & frequency, & Shift $\Delta_x$, mas  & Shift $\Delta_y$, mas& Beam,               & Peak flux,  \\
           & cies         & GHz        &                        &               & mas$\times$mas,$^\circ$] & Jy/beam  \\\hline
J2202+4216 & S+X      &  5.469   & $-0.185$ (S--X) & $-2.300$
(S--X)  & $1.58\times 1.00,  -14^\circ$ & 0.77    \\\hline
J0336+3218 & S+X      &  5.474   & $1.230$   (S--X) & $-0.680$
(S--X)  & $1.04\times 0.67, -13^\circ$ & 0.61    \\\hline
J1419+5423 & C+X+U    &  8.405   & $0.190$   (C--U) & $-0.170$ (C--U)  & $0.70\times 0.62,  75^\circ$ & 0.50    \\
           &          &          & $0.060$   (X--U) & $-0.050$ (X--U)  &                              &         \\\hline
\end{tabular}\end{small}
\end{center}

\begin{figure}[b]
{\begin{center}
 \includegraphics[width= 160mm]{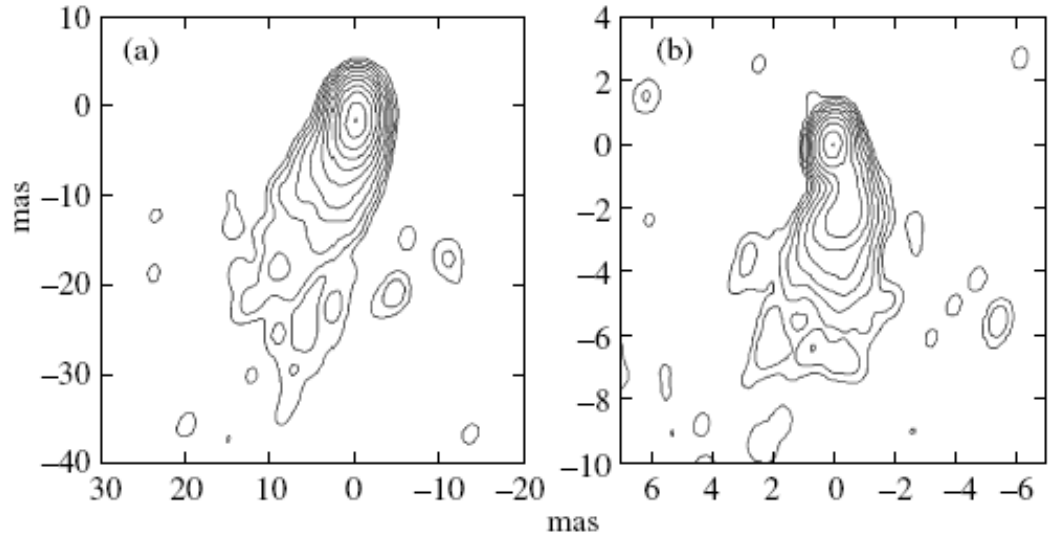}
 \caption{Intensity
maps for the radio source J2202+4216 at frequencies of (a) 2.3 and
(b) 8.6 GHz. The lower level of the contour lines is (a) 0.125\%
and (b) 0.25\% of the peak flux density (the peak flux densities
are listed in Table 1); the values of the succeeding levels are
doubled.}
\end{center}}
\end{figure}

\begin{figure}[b]
{\begin{center}
 \includegraphics[width= 160mm]{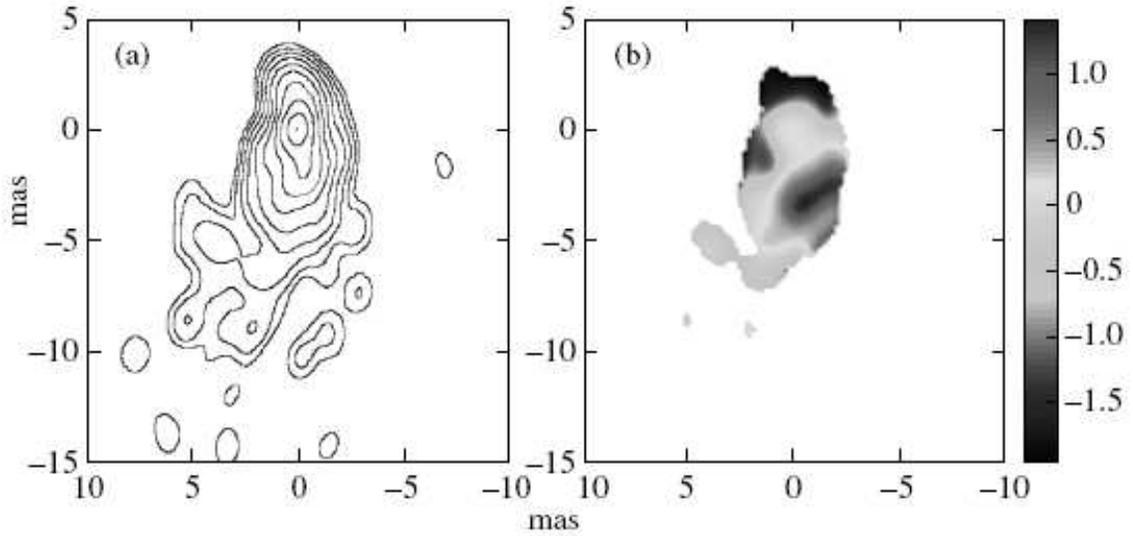}
 \caption{Results of two-frequency image synthesis for the radio
source J2202+4216 without any allowance for the mutual shift of
its single-frequency images: (a) intensity map (the lower level of
the contour lines is 0.25\% of the peak flux density, the values
of the succeeding levels are doubled), the reference frequency is
5.47 GHz; (b) spectral-index map.}
\end{center}}
\end{figure}

\begin{figure}[b]
{\begin{center}
\includegraphics[width= 160mm]{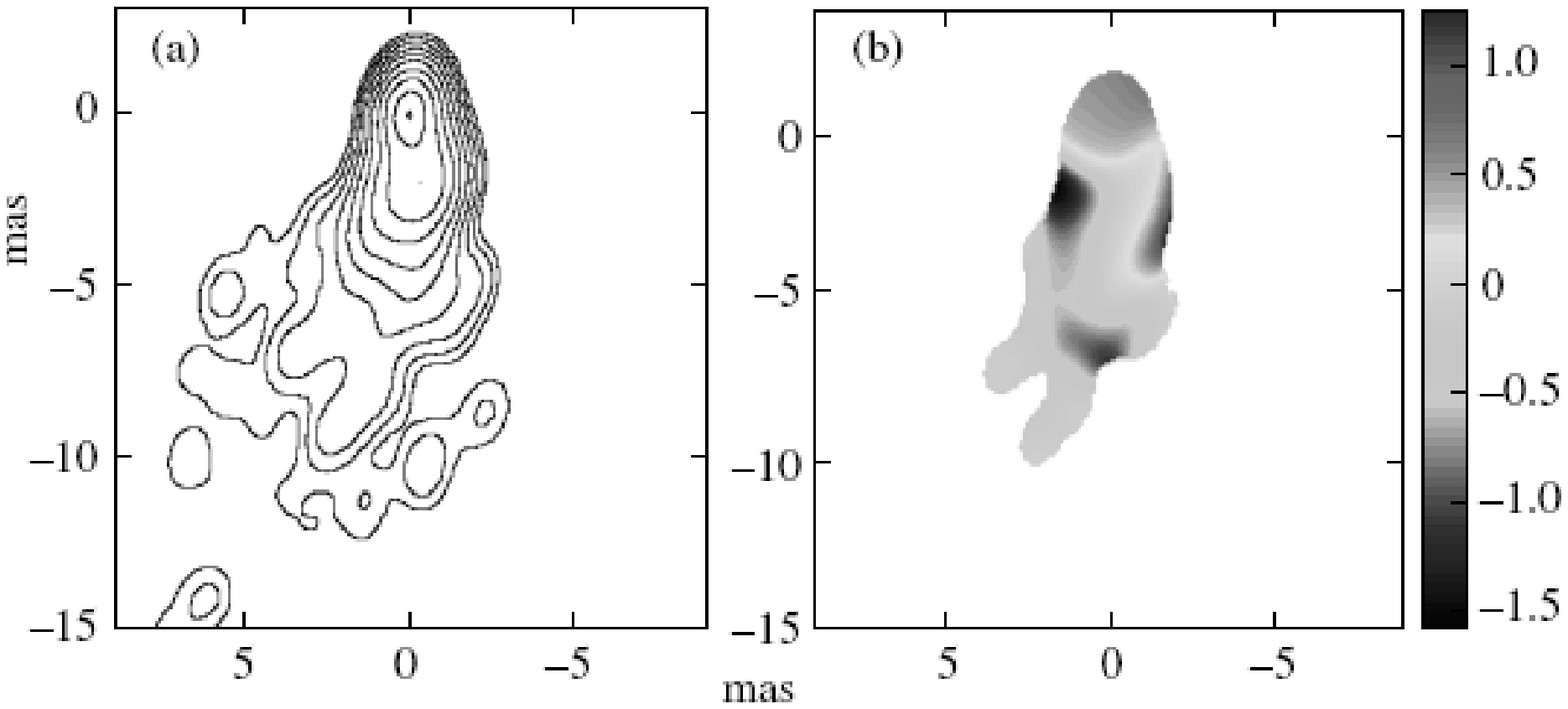}
\caption{Results of two-frequency image synthesis for the radio
source J2202+4216 with the pre-alignment of the peak values of its
single-frequency images: (a) intensity map (the lower level of the
contour lines is 0.25\% of the peak flux density, the values of
the succeeding levels are doubled); (b) spectral-index map.}
\end{center}}
\end{figure}

\begin{figure}[b]
{\begin{center}
\includegraphics[width= 160mm]{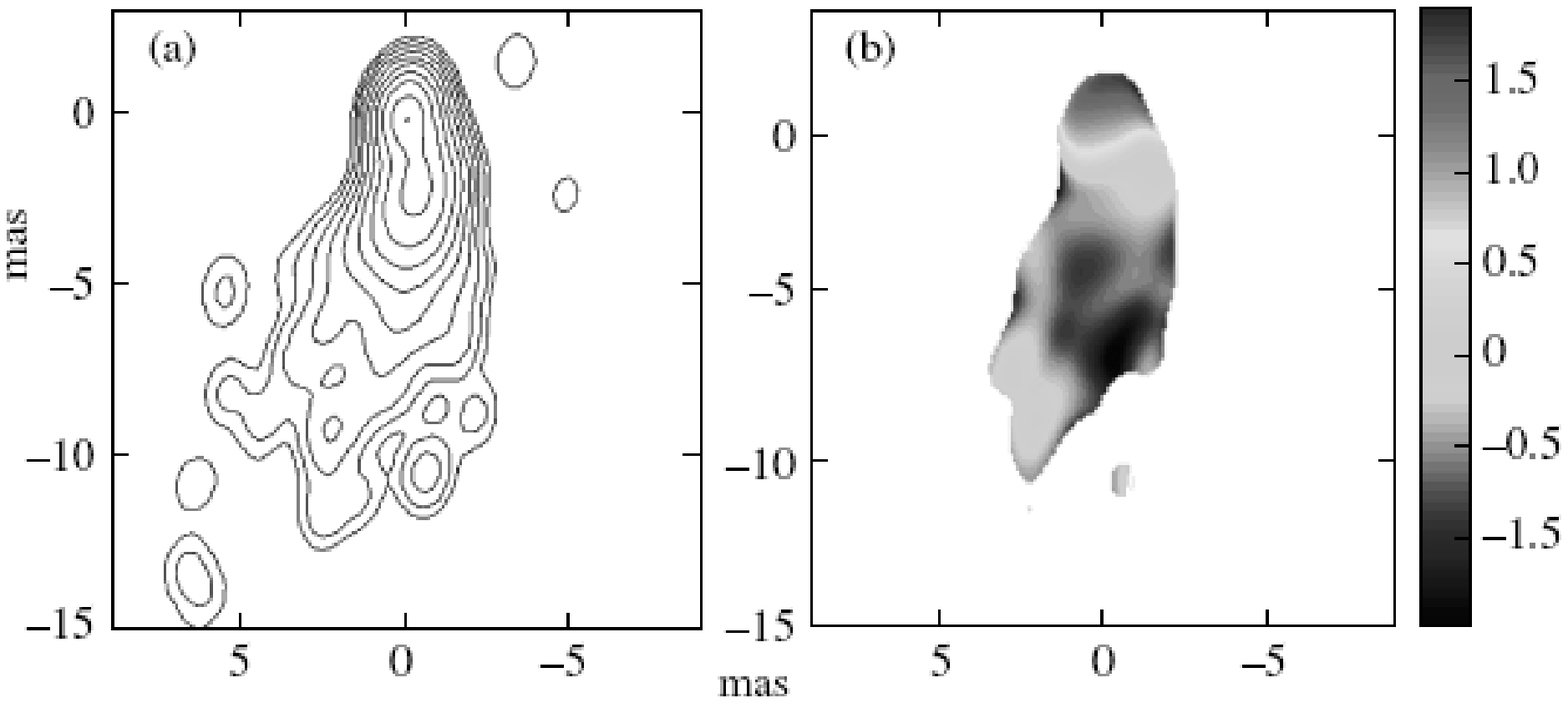}
\caption{Results of two-frequency image synthesis for the radio
source J2202+4216 with the pre-alignment of optically thin jet
features of its single-frequency images: (a) intensity map (the
lower level of the contour lines is 0.25\% of the peak flux
density (see Table 2), the values of the succeeding levels are
doubled); (b) spectral-index map.}
\end{center}}
\end{figure}

\begin{figure}[b]
{\begin{center}
\includegraphics[width= 160mm]{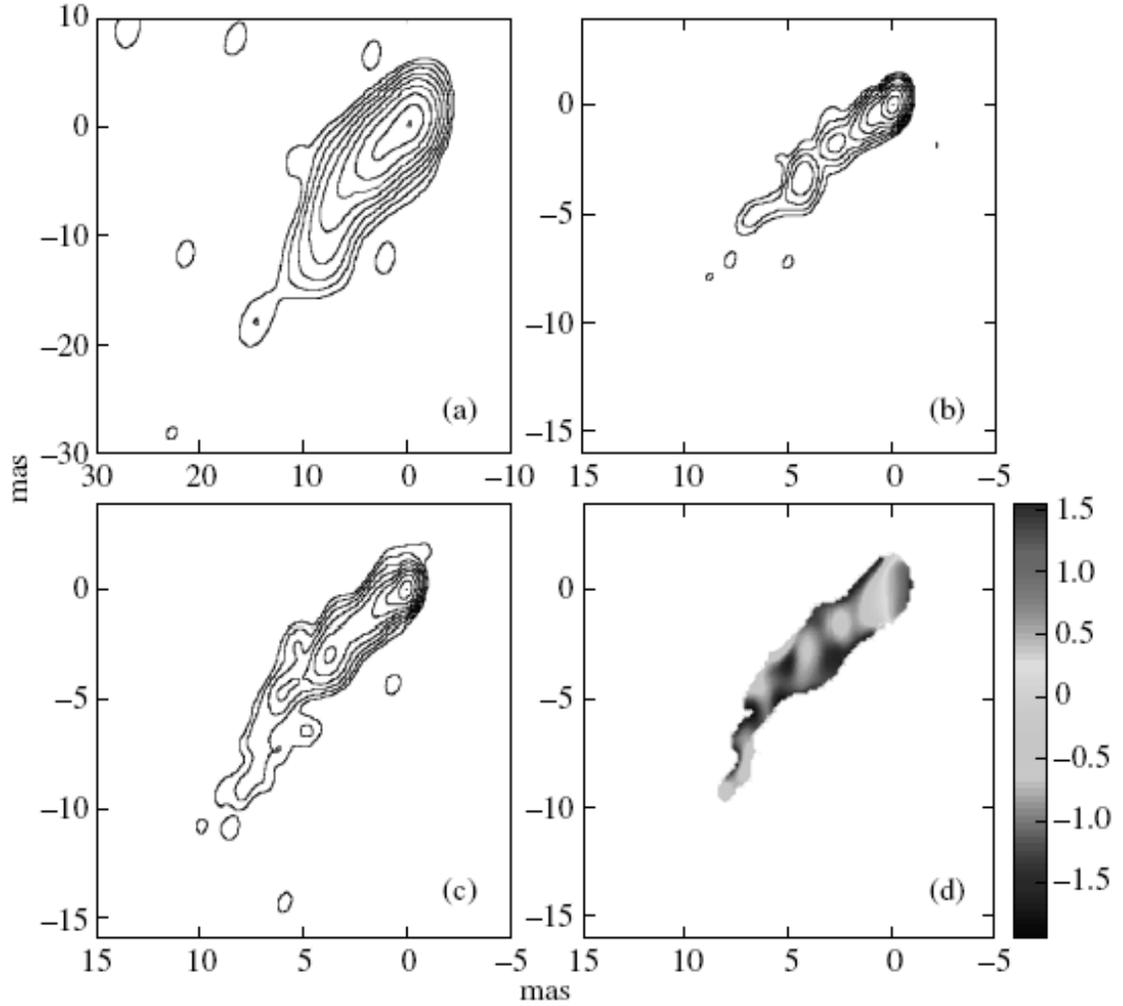}
\caption{Mapping results for the radio source J0336+3218: (a)
intensity map at 2.3 GHz (the lower level of the contour lines is
0.5\% of the peak flux density on the map, the values of the
succeeding levels are doubled); (b) intensity map at 8.6 GHz (the
lower level of the contour lines is 0.25\%) (the peak fluxes are
listed in Table 1); (c) two-frequency intensity map (the reference
frequency is 5.47 GHz) synthesized by taking into account the
mutual shift of the single-frequency images found by aligning
optically thin jet features (the lower level of the contour lines
is 0.25\%) (the peak flux is given in Table 2); (d) spectral-index
map.}
\end{center}}
\end{figure}

\newpage

\begin{figure}[b]
{\begin{center}
\includegraphics[width= 170mm]{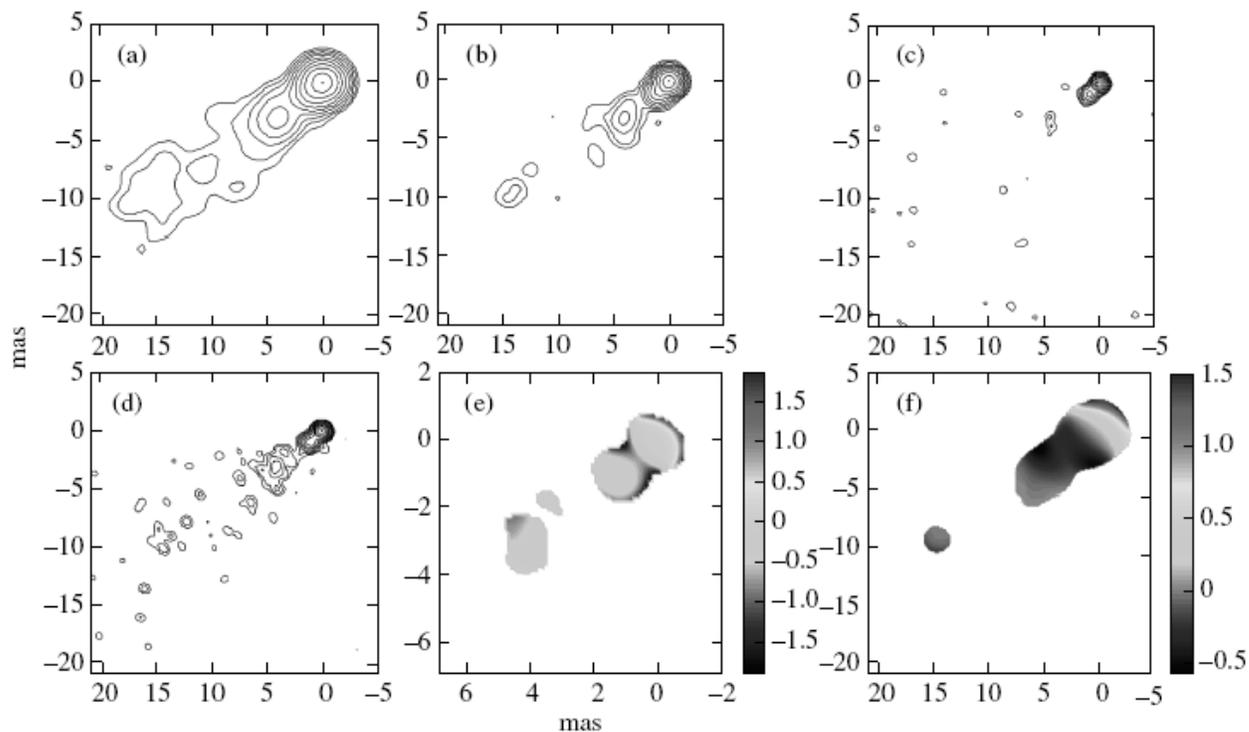}
\caption{Mapping results for the radio source J1419+5423:
intensity map at (a) 5, (b) 8.4, and (c) 15.3 GHz (the lower level
of the contour lines is 0.25\% of the peak flux density on the
map, the values of the succeeding levels are doubled) (the peak
fluxes are given in Table 1); (d) three-frequency intensity map
(the reference frequency is 8.4 GHz) synthesized by taking into
account the mutual shift of the single-frequency images found by
aligning optically thin jet features (the lower level of the
contour lines is 0.125\%)(the peak flux is given in Table 2); (e)
and (f) spectral-index maps with the resolutions corresponding to
the data in the U and C bands, respectively.}
\end{center}}
\end{figure}

\end{document}